\documentclass[journal]{IEEEtran}
\usepackage{amsfonts,amssymb,amsmath,bm}
\usepackage{float}
\usepackage{graphicx,subfigure}
\usepackage{cite,color}
\usepackage{cases}
\usepackage{epsfig}
\usepackage{verbatim}
\usepackage{algorithm}
\usepackage{dblfloatfix}
\usepackage{setspace}
\usepackage{tabularx}
\IEEEoverridecommandlockouts

\newtheorem{lemma}{\textbf{Lemma}}

\newtheorem{remark}{\textbf{Remark}}[section]

\ifCLASSINFOpdf
\else
\fi
\begin{document}
	%
	%
	%
	%
	
	\title{Energy Efficiency Optimization for Millimeter Wave System with
		Resolution-Adaptive ADCs}
	\author
	{
		Hualian Sheng, Xihan Chen, Xiongfei Zhai, An Liu, and Min-Jian Zhao
\thanks{
Hualian Sheng, Xihan Chen, An Liu and Min-Jian Zhao are with the College of Information
Science and Electronic Engineering, Zhejiang University, Hangzhou
310027, China (e-mail: hlsheng; chenxihan; anliu; mjzhao@zju.edu.cn).

Xiongfei Zhai is with  the School of Information Engineering, Guangdong University of Technology, Guangzhou 510006, China
(e-mail: feifei4006@163.com).}
%
	}

	\linespread{0.9}  
	\maketitle
	\pagestyle{empty}
	\thispagestyle{empty}
	\begin{abstract}
This letter investigates the uplink of a multi-user millimeter wave (mmWave) system, where the base station (BS) is equipped with a massive multiple-input multiple-output (MIMO) array and resolution-adaptive analog-to-digital converters  (RADCs). Although employing massive MIMO at the BS can significantly improve the spectral efficiency, it also leads to high hardware complexity and huge power consumption.
To overcome these challenges, we seek to jointly optimize the beamspace hybrid combiner and the ADC quantization bits allocation to maximize the system energy efficiency (EE) under some practical constraints.
The formulated problem is non-convex due to the non-linear fractional objective function and the non-convex feasible set which is generally intractable. In order to handle these difficulties, we first apply some fractional programming (FP) techniques and introduce auxiliary variables to recast this problem into an equivalent form amenable to optimization.
Then, we propose an efficient double-loop iterative algorithm based on the penalty dual decomposition (PDD) and the majorization-minimization (MM) methods to find local stationary solutions. Simulation results reveal significant gain over the baselines.
\end{abstract}
	
	\IEEEpeerreviewmaketitle
	\section{Introduction}
	Millimeter wave (mmWave) communication has become a key enabling technology to accommodate the ever increasing data traffic in fifth generation (5G) systems. The shorter  wavelength of antenna components to be packed into physical devices with small trade-off \cite{ladc_2015}, which enables large spatial multiplexing and highly directional combining. Nevertheless, the traditional fully-digital combining scheme requires dedicated radio frequency (RF) chain with power-demanding high-resolution analog-to-digital converters  (ADCs) per antenna element, which leads to huge hardware cost and power consumption at the base station (BS).

	To address these limitations, the selection/optimization on the number of RF chains is a potential solution for reducing power consumption and hardware complexity \cite{Kaushik_2019}. Furthermore, the convergence of hybrid combining and low-resolution ADCs (LADCs) is becoming an evident trend for future wireless network and has drawn considerable academic interests in recent years. The authors of  \cite{JMOfew_adc_2016} characterized the trade-off between the achievable rate and power consumption in the hybrid combiner architecture with LADCs.  By exploiting the sparse nature of mmWave channels, the authors of \cite{WHAO_arxiv_2019} invoked the beamspace massive multiple-input multiple-output (MIMO) techniques to steer the arriving signals with different angle of arrivals  to distinct array elements, which significantly reduces the number of RF chains required and further achieves cost-effective implementations.
However, the nonlinear distortion caused by LADCs would inevitably lead to huge performance degradation in the high signal-to-noise ratio (SNR) regime.
To overcome this deficiency, the authors of \cite{mixed_adc} suggested implementing a mixed-ADC architecture, in which a combination of low- and high-resolution ADCs are used side-by-side. To achieve more preferable energy-rate trade-off, the authors of \cite{JCHOI_BA_2017} proposed the hybrid MIMO receiver architecture with resolution-adaptive ADCs (RADCs). In addition, two heuristic ADC quantization bits allocation algorithms were conceived to minimize the total quantization error under the total ADC power consumption constraint. 


The joint optimization has been considered for point-to-point communication systems \cite{AKau_EE_pp}. However, for the multi-user uplink communication system, the existing works in \cite{WHAO_arxiv_2019, mixed_adc, JCHOI_BA_2017} only consider separate optimization. Besides, all the aforementioned schemes are based on heuristic or separate optimization of beamspace hybrid combining and ADC quantization bits allocation, which might suffer from significant performance degradation. There is scope for further research on energy efficiency (EE) maximization for the beamspace massive MIMO architecture with RADCs, despite its paramount importance to practical implementation and performance improvement.

Contribution of this letter includes the algorithm design for joint beamspace hybrid combining and ADC quantization bits
allocation (JBQA) scheme for the uplink transmission of mmWave systems with RADCs, to maximize the system EE. In particular, the resolution of each RADC can be dynamically adjusted to mitigate the quantization error according to the channel gain on the corresponding RF chain, leading in turn to reduced power consumption and improved system throughput.
We propose a \emph{fractional majorization-minimization penalty dual decomposition} (FMP) algorithm to solve this joint optimization problem.
Simulation results verify the advantages of the proposed JBQA scheme over the state-of-the-art baselines.
	\section{System Model and Problem Formulation}
	\subsection{Network Architecture}
	This paper considers a multi-user beamspace mmWave uplink
	system, where the BS serves $K$ single-antenna users by using a massive array of $N$ antennas and $M$ RF chains. We consider an extended Saleh-Valenzuela geometric model for mmWave channels \cite{channel_s}. Furthermore, the channel is considered as block flat-fading due to the small delay spread of mmWave channel. The channel between the BS and the $k$-th user is denoted by $\bm{h}_k\in \mathbb{C}^{N}$, and the received signal at the BS can be expressed as

	\begin{align}
	\bm{y}=\sum_{k=1}^K \sqrt{p}_k \bm{h}_k s_k+{\bm{n}}=\mathbf{H}
	\mathbf{P}\bm{s}+{\bm{n}},
	\end{align}
	where $\mathbf{H}\triangleq[\bm{h}_1,\cdots,\bm{h}_K]$, $\mathbf{P}\triangleq\mathrm{diag}(\sqrt{p_1},\cdots,\sqrt{p_K})$ with $p_k$ being the
	transmit power of user $k$, $\bm{s}\triangleq[s_1,\cdots,s_K]^T$ with $s_k\sim
	\mathcal{CN}(0,1)$ being the data symbol of user $k$, and ${\bm{n}}\sim
	\mathcal{CN}(0,\sigma^2_w \mathbf{I}_N)$ is the additive white Gaussian noise.


	\setcounter{equation}{13}
	\begin{figure*}[b]
		\hrulefill
		\begin{align}
		\hat{r}_k(\bm{z},{\phi}_k,{\lambda}_k)&=\log(1+\phi_k)-\phi_k
		+2\mathfrak{R}e\{\sqrt{p_k(1+\phi_k)}\lambda^H_k\bm{u}^H_k\mathbf{D}^H\mathbf{F}_{\alpha}\mathbf{Q}^H
		\bm{h}_k\}-\lambda^H_k\lambda_k \gamma_k(\bm{z}),\label{topE2_1}\\
		\gamma_k(\bm{z})&=\sum_{l=1}^K p_l
		|\bm{u}^H_k\mathbf{D}^H\mathbf{F}_{\alpha}\mathbf{Q}^H \bm{h}_l |^2+\sigma^2_w
		\bm{u}^H_k\mathbf{D}^H\mathbf{F}_{\alpha}\mathbf{Q}^H
		\mathbf{Q}\mathbf{F}_{\alpha}\mathbf{D}\bm{u}_k+
		\bm{u}^H_k\mathbf{D}^H \mathbf{A}_q \mathbf{D}\bm{u}_k.\label{topE2_2}
		\end{align}
	\end{figure*}
	\setcounter{equation}{2}
\subsection{Proposed JBQA Scheme}	
In this paper, we devise a JBQA scheme to achieve large spatial multiplexing and array gain while addressing the hardware limitations. The received signal $\bm{y}$ is processed by a beamspace hybrid combiner with RADCs. In this case, the received signal at the BS is first combined via the beamspace RF combiner $\mathbf{Q}\triangleq\mathbf{WG}\in \mathbb{C}^{N\times M}$, where $\mathbf{W}\in \mathbb{C}^{N\times S}$ is the codebook of size $S$ and $\mathbf{G}\in  \mathbb{C}^{S\times M}$ is the selection matrix  with binary entry $g_{sm}\in\{0,1\}$ to choose codewords. Using the above notations, the  signal after the beamspace RF combiner can be expressed as
	\begin{equation}\label{RF_y}
	\hat{\bm{y}}=\mathbf{Q}^H\mathbf{H}\mathbf{P}\bm{s}+\mathbf{Q}^H{\bm{n}},
	\end{equation}
	
 	We consider that $M$ pairs of RADCs are connected to RF processors, enabling more flexible and refined control on quantization bits allocation to mitigate the quantization error. Assuming that the coefficients of automatic gain control (AGC) are appropriately set, the linear additive quantization noise model (AQNM) is adopted to approximate the quantization process \cite{ADMM}, where each RADC quantizes either real or imaginary component of $\hat{\bm{y}}$. The resulting signal is given by
	\begin{align}
	\tilde{\bm{y}}=\mathcal{F}(\hat{\bm{y}})=\mathbf{F}_{\alpha}\hat{\bm{y}}+\bm{n}_q,
	\end{align}
	where $\mathcal{F}(\cdot)$ is the element-wise quantizer operator;
	$\mathbf{F}_{\alpha}\triangleq\mathrm{diag}(\alpha_1,\cdots,\alpha_{M})\in
	\mathbb{C}^{M\times M}$ is the quantization
	gain matrix with $\alpha_m\triangleq1-\beta_m$ , where $\beta_m\triangleq \frac{\pi \sqrt{3}}{2} 4^{-b_m}$ is the
	normalized quantization error when the number of quantization bits is $b_m$;
	$\bm{n}_q$ is the additive quantized noise distributed with zero mean and the
	covariance matrix
	$
	\mathbf{A}_{q}\triangleq\mathbf{F}_{\alpha}\mathbf{F}_{\beta}\mathrm{diag}(
	\mathbf{Q}^H\mathbf{H}\mathbf{P}^2\mathbf{H}^H\mathbf{Q}
	+\sigma^2_w \mathbf{Q}^H\mathbf{Q})
	$
	with $\mathbf{F}_{\beta}\triangleq\mathrm{diag}(\beta_1,\cdots,\beta_{M})\in
	\mathbb{C}^{M\times M}$. Note that $\bm{n}_q$ is uncorrelated with $\hat{\bm{y}}$ \cite{ADMM}.
	
	Then, the quantized signal is combined by the baseband combiner $\mathbf{D} \in \mathbb{C}^{M \times M}$ to reduce the quantization loss and combat the multi-user interference. Finally, the combined signal is detected by linear receiver $\bm{u}_k \in \mathbb{C}^{M}$.
	Based on the above procedure, the detected signal for user $k$ is given by
	
	\begin{align}
	\hat{s}_k
	\!=\!\bm{u}^H_k \mathbf{D}^H\mathbf{F}_{\alpha}\mathbf{Q}^H\mathbf{H}
	\mathbf{P}^{\frac{1}{2}}\bm{s}
	+\bm{u}^H_k \mathbf{D}^H\mathbf{F}_{\alpha}\mathbf{Q}^H\hat{\bm{n}} +\bm{u}^H_k
	\mathbf{D}^H\bm{n}_q.\nonumber
	\end{align}
	\subsection{Achievable Rate and Power Consumption}
	For convenience, we define $\bm{p}\triangleq[p_1,\cdots,p_K]^T$,
	$\bm{g}\triangleq\mathrm{vec}(\mathbf{G})$, $\bm{d}\triangleq\mathrm{vec}(\mathbf{D})$,
	$\bm{u}\triangleq[\bm{u}^T_1,\cdots,\bm{u}^T_K]^T$, $\bm{b}\triangleq[b_1,\cdots,b_{M}]^T$, and
	$\bm{z}\triangleq[\bm{p}^T,\bm{g}^T,\bm{d}^T,\bm{u}^T,\bm{b}^T]^T$. Using the
	above notations, the achievable data rate of user $k$ is
	\begin{align}
	r_k(\bm{z})=\log(1+\theta_k(\bm{z})),
	\end{align}
	where
	$\theta_k(\bm{z}) \triangleq {\theta_k^{\alpha}(\bm{z})}/{\theta_k^{\beta}(\bm{z})}$ is the signal-to-interference-plus-noise ratio (SINR), $\theta_k^{\alpha}(\bm{z})\triangleq p_k
	|\bm{u}^H_k\mathbf{D}^H\mathbf{F}_{\alpha}\mathbf{Q}^H \bm{h}_k |^2$ is the desired signal term, and $\theta_k^{\alpha}(\bm{z}) \triangleq \sum_{l\neq k}p_l |\bm{u}^H_k\mathbf{D}^H\mathbf{F}_{\alpha}\mathbf{Q}^H
	\bm{h}_l |^2+\sigma^2_w \bm{u}^H_k\mathbf{D}^H\mathbf{F}_{\alpha}\mathbf{Q}^H
	\mathbf{Q}\mathbf{F}_{\alpha}\mathbf{D}\bm{u}_k+
	\bm{u}^H_k\mathbf{D}^H \mathbf{A}_q \mathbf{D}\bm{u}_k$ is the interference-plus-noise term.
	
	Moreover, the system power consumption is given by
	\begin{align}
	P_C(\bm{z})=\sum_{k=1}^K p_k+ \sum_{m=1}^{M} p_m^A(b_m) +P_o,
	\end{align}
	where $\sum_{k=1}^K p_k$ is the total power consumed by users (i.e., the transmit power); $
	P_o\triangleq P_B+M(P_{R}+P_S+P_L)
	$, and $P_B$ is the power consumption of the baseband combiner, $P_R$ is the
	power consumed per RF chain, $P_S$ is the power consumed per switch, $P_L$ is
	the power consumed per low noise amplifier;
	$
	p_m^A(b_m) \triangleq \varphi f_s2^{b_m+1}
	$ is the power consumption of the $m$-th pair of RADCs, where $\varphi$ is
	the power consumed per conversion procedure and $f_s$ is the Nyquist
	sampling rate.

	\subsection{Problem Formulation}
	Due to the ever-increasing number of mobile users, EE in the mmWave uplink is of high priority as user terminals are power-constrained \cite{Ming_Zeng}. Our interest in this letter lies in the joint optimization of beamspace hybrid combiner and ADC quantization bits allocation to maximize the system EE. The resulting problem can be formulated as
	\begin{align}\label{P_1}
	&\mathop{\max}_{\bm{z}}~ \frac{\sum_{k=1}^K
		r_k(\bm{z})}{P_C(\bm{z})},\\
	&~\textrm{s.t.} \quad 0 \leq p_k \leq P_k^{\mathrm{max}}, \forall k \label{p_limit_1},\\
	&~\quad\quad \sum_{s=1}^S g_{sm}=1, \sum_{m=1}^{M} g_{sm}\leq1,
	g_{sm}\in\{0,1\}\label{b_limit_1},\forall s,m,\\
	&~\quad\quad b_m^{\mathrm{min}} \leq b_m \leq b_m^{\mathrm{max}} \text{\ is an integer}, \forall
	m,\label{d_limit_1}\\
	&~\quad\quad \sum_{m=1}^{M}{b_m} \leq {M}\bar{b},\label{d_limit_2}
	\end{align}
where $P_k^{\mathrm{max}}$ is the transmit power budget for user $k$, $b_m^{\mathrm{min}}$ and $b_m^{\mathrm{max}}$ respectively are the minimum and maximum of the quantization bits, and $\bar{b}$ is the average quantization bits. Constraint \eqref{b_limit_1} is added to guarantee that each RF chain is associated with only one codeword, and each codeword is selected for at the most one RF chain. Constraint \eqref{d_limit_1} represents the limitations on the quantization bits for each RADC. Constraint \eqref{d_limit_2} gives a reference total ADC quantization bits for the above EE optimization problem.

	\setcounter{equation}{11}

Solving problem \eqref{P_1} is  difficult due to the following reasons. First, both the selection matrix $\mathbf{G}$ and quantization bits vector $\bm{b}$ are discrete, which makes the feasible set non-convex. Second,  the optimization variables are highly coupled in the non-convex objective function and constraints. In a nutshell, we are faced with a mixed-integer nonlinear programming (MINP) problem, which is usually considered as NP-hard.
	

\section{Proposed FMP Algorithm}
In thus section, we first transform the original problem \eqref{P_1} into a more tractable yet equivalent form by exploiting some fractional programming (FP) techniques \cite{Dinkelbach, FP_2}. Subsequently, we develop an efficient double-loop iterative algorithm based on majorization-minimization (MM) \cite{MM_method} and penalty dual decomposition (PDD) methods  \cite{PDD} to find its local stationary solutions.
	\subsection{Problem Reformulation}
	With the aid of Dinkelbach method \cite{Dinkelbach}, we can transform problem \eqref{P_1} into a more tractable yet equivalent form. 
	\begin{lemma}\label{lemma1}
		\emph{Let $\mathcal{Z}=\{z|\eqref{p_limit_1}-\eqref{d_limit_2}\}$ denote the feasible set of $z$ in problem (7). Then, it is equivalent to the following}
		\begin{align}\label{P_2}
		\max_{\bm{z}\in \mathcal{Z},\eta}\: \sum_{k=1}^K r_k(\bm{z})-\eta P_C(\bm{z})
		\end{align}
		\emph{where $\eta^{\star}=\frac{\sum_{k=1}^K
				r_k(\bm{z}^{\star})}{P_C(\bm{z}^{\star})}$ is the optimal trade-off between the sum
			rate and the power consumption.}
	\end{lemma}

We remark that $\sum_{k=1}^K r_k(\bm{z})$ is in the sum-ratio form, where each ratio term is embedded in the log function. To tackle these difficulties, we adopt Lagrangian
	dual transform and complex quadratic transform methods \cite{FP_2} to reformulate
	problem (\ref{P_2}).
	\begin{lemma}\label{lemma2}
		\emph{The optimal $\bm{z}^{\star}$ is solved if and only if it solves}
		\begin{align}\label{P_3}
		\max_{\bm{z}\in \mathcal{Z},\eta,\bm{\phi},\bm{\lambda}}\: \sum_{k=1}^K
		\hat{r}_k(\bm{z},{\phi}_k,{\lambda}_k)-\eta P_C(\bm{z})
		\end{align}
		\emph{where $\hat{r}_k(\bm{z},{\phi}_k,{\lambda}_k)$ with $\gamma_k(\bm{z})$ is
			defined in \eqref{topE2_1}-\eqref{topE2_2} as displayed at the bottom of this page,
			$\lambda^{\star}_k\triangleq\sqrt{p_k(1+\phi_k)}\bm{u}^H_k\mathbf{D}^H\mathbf{F}_{\alpha}\mathbf{Q}^H
			\bm{h}_k/\gamma_k(\bm{z})$ is the auxiliary variable introduced for taking ratio terms out of log function, and $\bm{\phi}^{\star}\triangleq\theta_k(\bm{z})$ is the vector of auxiliary
			variable introduced for linearizing each ratio term in $r_k$.}
	\end{lemma}
	\setcounter{equation}{15}
	
It should be noteworthy that the numerator and denominator in problem  \eqref{P_2} are now decoupled in the reformulated problem \eqref{P_3}.  Furthermore, we relax discrete constraint \eqref{d_limit_1} into a closed connected subset of the real axis, i.e.,
	\begin{align}\label{d_limit_12}
	b_m^{\mathrm{min}} \leq b_m \leq b_m^{\mathrm{max}}, \forall m,
	\end{align}

Similar to \cite{round_bit}, we round each $b_m$ as follows
	\begin{equation}
	\bar{b}_m(\delta)=\left\{
	\begin{aligned}
	\overset{} \lfloor{b_m^{\star}}\rfloor&,
	~~~~~\text{if}~b_m^{\star}-\lfloor{b_m^{\star}}\rfloor \leq \delta  \\
	\lceil{b_m^{\star}}\rceil&,~~~~~\text{otherwise,}\\
	\end{aligned}
	\right.\forall m,
	\end{equation}
	where $0 \leq\delta\leq 1$ is chosen such that $\sum_{m=1}^{M}\bar{b}_m(\delta)
	\leq {M}\bar{b}$.
	
To overcome the difficulty posed by  discrete binary constraint \eqref{b_limit_1}, a suitable transformation is necessary. To this end, we rewrite constraint \eqref{b_limit_1} into the following equivalent form:
	\begin{align}
	&\bm{g}^T_s \bm{e}_m=\hat{g}_{sm},~\sum_{s=1}^S \bm{g}^T_s \bm{e}_m=1,
	~\bm{g}^T_s \bm{e}_m(1-\hat{g}_{sm})=0,\label{constraintb_21}\\
	&\bm{g}^T_s \bm{1}_{M}\leq 1,~~~0\leq \hat{g}_{sm} \leq 1,\label{constraintb_22}
	\end{align}
where $\bm{g}_s\triangleq[g_{s1},\cdots,g_{sM}]^T$ is the $s$-th row of $\mathbf{G}$, $\bm{e}_m$ is the $m$-th column of identity matrix $\mathbf{I}$, and
$\bm{1}_{M}\triangleq[1,\cdots,1]^T$.

For clarity, we define $\hat{\mathcal{Z}}=\{\hat{\bm{z}}\triangleq
	[\bm{z}^T,\hat{\bm{g}}^T]^T|\eqref{p_limit_1},\eqref{d_limit_2},\eqref{d_limit_12}-\eqref{constraintb_22}\}$ with
	$\hat{\bm{g}}\triangleq[\hat{\bm{g}}^T_1,\cdots,\hat{\bm{g}}^T_S]^T$, and
	$\hat{\bm{g}}_s\triangleq[\hat{g}_{s1},\cdots,\hat{g}_{sM}]^T$.
By penalizing and dualizing constraint \eqref{constraintb_21} into objective function \eqref{P_3}, we obtain the following augmented Lagrangian problem:
	\begin{align} &\mathop{\max}_{\hat{\bm{z}}\in
		\mathcal{Z},\eta,\bm{\phi},\bm{\lambda}}~
	\mathcal{J}(\hat{\bm{z}},\eta,\bm{\phi},\bm{\lambda}),\label{AL_obj}
	\end{align}
	where
	\begin{align}
		&\notag\mathcal{J}(\hat{\bm{z}},\eta,\bm{\phi},\bm{\lambda})=\sum_{k=1}^K
		\hat{r}_k(\bm{z},{\phi}_k,{\lambda}_k)-\eta
		P_C(\bm{z})-\frac{1}{2\rho}\Bigg (\\\notag
		&\sum_{s=1}^{S}\!\sum_{m=1}^{M}\!\!\left((\bm{g}^T_s
		\bm{e}_m\!-\!\hat{g}_{sm}\!+\rho\zeta_{sm})^2 \!+\! (\bm{g}^T_s
		\bm{e}_m(1\!-\!\hat{g}_{sm})+\rho\nu_{sm})^2\right)\\\notag
		&+\sum_{m=1}^{M}(\sum_{s=1}^S \bm{g}^T_s \bm{e}_m-1+\rho\varsigma_m)^2\Bigg ),
	\end{align}
$\{\zeta_{sm}\}$, $\{\varsigma_m\}$, $\{\nu_{sm}\}$ are the Lagrange multipliers, and $\rho$ is penalty coefficient.

\subsection{The proposed FMP Algorithm}
	In this subsection, we elaborate the implementation details of the proposed FMP algorithm which exhibits a double-loop structure: 1) optimization variables are updated in the inner loop by iteratively solving problem \eqref{AL_obj}; 2)  dual variables and penalty parameter are updated based on the constraint violation indicator in the outer loop. Hereinafter, we introduce $t$ as outer iteration index and $v$ as the inner iteration index.
	In particular, the dual variables are updated as follows
	\begin{subequations}\label{dualVar}
		\begin{align}
		\zeta^{t+1}_{sm}&=\zeta^{t}_{sm}+(\bm{g}^T_s \bm{e}_m-\hat{g}_{sm})/{\rho^t},\\
		\varsigma^{t+1}_m&=\varsigma^t_m+(\sum_{s=1}^N\bm{g}^T_s \bm{e}_m-1)/{\rho^t},\\
		\nu^{t+1}_{sm}&=\nu^{t}_{sm}+\bm{g}^T_s \bm{e}_m(1-\hat{g}_{sm})/{\rho^t},
		\end{align}
	\end{subequations}
	The constraint violation indicator $\epsilon(\hat{\bm{z}})$ is given by
	\begin{equation}
	\epsilon(\hat{\bm{z}})=\max \limits_{s,m}\{|\bm{g}^T_s
	\bm{e}_m-\hat{g}_{sm}|,|\sum_{s=1}^S\bm{g}^T_s \bm{e}_m-1|,|\bm{g}^T_s
	\bm{e}_m(1-\hat{g}_{sm})|\}.\nonumber
	\end{equation}

Observe that constraints in problem \eqref{AL_obj} are separable, so it allows us to decompose the original problem into nine independent blocks.
The corresponding subproblem for each block can be efficiently solved with the others fixed. Given the penalty parameter $\rho$ and the dual variables $\{\zeta_{sm}, \varsigma_{m}, \nu_{sm}\}$, the details of the inner iteration are elaborated below.

	\subsubsection{Optimization of $\bm{p}$}
	It shows that subproblem w.r.t. $\bm{p}$ is a linearly constrained concave optimization problem, which can be efficiently solved by the Lagrangian multiplier
	method \cite{Language}. Consequently, the optimal $\bm{p}^{\star}$ is given by
	\begin{equation}\label{updateP}
	p^{\star}_k(\sigma_k)=(\mathfrak{R}e\{\sqrt{1+\phi_k}\lambda^H_k\bm{u}^H_k\mathbf{D}^H\mathbf{F}_{\alpha}\mathbf{Q}^H
	\bm{h}_k\}/\vartheta_k(\sigma_k))^2,
	\end{equation}
	where
	$
	\vartheta_k(\sigma_k)\triangleq\sigma_k+\eta+\sum_{l=1}^K
	|\lambda^H_l\bm{u}^H_l\mathbf{D}^H\mathbf{F}_{\alpha}\mathbf{Q}^H \bm{h}_k|^2
	+\sum_{l=1}^K\sum_{m=1}^{M} |\lambda_l|^2 [\mathbf{F}_{\alpha}\mathbf{F}_{\beta}]_m
	|[\mathbf{Q}^H\mathbf{H}]_{mk}|^2|[\mathbf{D}\bm{u}_l]_m|^2
	$, $\sigma_k$ is chosen to be zero if $p_k(0)\leq
	P^{\mathrm{max}}_k$ and chosen to satisfy $p_k(\sigma_k)= P^{\mathrm{max}}_k$
	otherwise. Note that $[\cdot]_m$ is the operator to take the $m$-th element of vector.
	$[\cdot]_{mk}$ is the operator to take the element in the $m$-th row and
	the $k$-th column of matrix.
	
\subsubsection{Optimization of $\bm{g}$, $\hat{\bm{g}}$, $\bm{u}$, $\bm{d}$}
All these subproblems w.r.t. $\bm{g}$, $\hat{\bm{g}}$, $\bm{u}$, $\bm{d}$ are unconstrained quadratic optimization problems, which can be solved by checking
the first-order optimality condition. The details are omitted due to the space limited.
	
\subsubsection{Optimization of $\bm{b}$}
Note that we cannot obtain the optimal $\bm{b}$ by directly maximizing the subproblem w.r.t.  $\bm{b}$ because the subproblem w.r.t. $\bm{b}$ is non-concave.
 By preserving the partial concavity of the original function and adding the proximal regularization, the concave surrogate function $g^v(\bm{b})$ for the $v$-th inner iteration is constructed as
	\begin{align}\label{MM_obj}
	g^v(\bm{b}) =
	\mathcal{J}(\bm{b}^v)+(\bm{\omega})^T(\bm{b}-\bm{b}^v)+\vartheta||\bm{b}-\bm{b}^v||^2,
	\end{align}
where $\vartheta < 0$ so that the surrogate function $g^v(\bm{b})$ is strongly concave, and $\bm{\omega}$ is the Jacobian
matrix of the objective function $\mathcal{J}(\hat{\bm{z}},\eta,\bm{\phi},\bm{\lambda})$ with respect to $\bm{b}$.
Then, the optimal solution $\bm{b}$ is obtained by solving the following strongly concave problem:
	\begin{align}\label{Update_d}
		\notag&\mathop{\max}_{\bm{b}}~ g^v(\bm{b}),\\
		&\textrm{s.t.} \quad b_m^{\mathrm{min}} \leq b_m \leq b_m^{\mathrm{max}}, \forall m,
	\end{align}
	which has no closed-form solution due to multiple simultaneous constraints. It can be efficiently solved by toolbox CVX.
	
	The proposed FMP algorithm is summarized in Algorithm \ref{alg:FMP}. Here, we emphasize that every limit point, denoted as $\bm{z}^{\star}$, generated by Algorithm 1 strictly satisfies the equality constraints (16) by adjusting the dual variables and penalty coefficient in a specific manner. Then we can show that  $\bm{z}^{\star}$ is a stationary point of problem \eqref{P_1}. The proof is similar to that of \cite{MM_method,PDD}, and it is hence omitted due to the limited space. Next, we compare the computational complexity of the proposed FMP algorithm with the following baseline schemes.
	\begin{itemize}
		\item \emph{FDC}: This scheme is implemented by fully digital combiner, aiming to maximize the system EE in \cite{FDP_scheme}.
		\item \emph{MMSQE-BA}: This is a variation of MMSQE-BA in \cite{JCHOI_BA_2017}  with the consideration of jointly optimizing the user power allocation and digital combiner.
	\end{itemize}
The overall computational complexity per iteration of the proposed JBQA scheme is $O(M^6 + K^2M^4 + M^3S + M^2N + KN^2)$ floating point operations (FPOs). The proposed JBQA scheme has much lower computational complexity than FDC scheme, which is $O(N^6)$ FPOs. Although the MMSQE-BA scheme provides a slightly lower computational complexity, which is $O(M^6 + K^2M^4 + M^2N + KN^2)$ FPOs, its performance is worse than that of our proposed JBQA scheme. Consequently, our proposed JBQA scheme offers a better trade-off between computational complexity and performance.

	
	\begin{remark}
		The proposed JBQA scheme can be easily extended to the case with multiple RF chains at the users. Specifically, we first transform the original problem into a more tractable yet equivalent form by some multi-dimensional FP techniques \cite{Dinkelbach}, \cite{FP_2} and subsequently develop an efficient algorithm based on matrix MM \cite{Sun_2016} and PDD \cite{PDD} methods.
	\end{remark}
	\begin{remark}
		The objective function (7) can be expanded by a hyper parameter $0 \leq \kappa \leq 1$ into the form $\max_{\bm{z} \in \mathcal{Z}}$$\sum_{k=1}^Kr_k(\bm{z})/(P_{T}({\bm{z}}))^{\kappa}$, where $\kappa$ represents the priority of EE as compared to the system throughput. Note that more emphasis is given to the maximization of the system throughput when $\kappa=0$. When $\kappa$ is relatively large, more weight is given to the power consumption minimization. In order to achieve a favorable trade-off between the system throughput and EE, it may require judicious selection for the $\kappa$. The resulting problem can be solved by the proposed FMP algorithm with minor modification. For the nonconcavity introduced by $\kappa$, it can be solved by the MM method.		
	\end{remark}

	\begin{algorithm}[H]
		\caption{\label{alg:FMP}Proposed FMP Algorithm for Problem \eqref{AL_obj}}
		
		\textbf{Initialization:}{\small{} $\hat{\bm{z}}^0 \in
			\hat{\mathcal{Z}}$, $\rho^0>0$, $\hat{\eta}$, $\{\zeta^0_{sm},\varsigma^0_m,\nu^0_{sm}\}$,
			$\mu^0>0$, $0<\chi<1$, accuracy tolerance $\tau$, the maximum inner iteration
			number $V$, the maximum outer iteration
			number $T$, $t=0$, $v=0$.}{\small\par}
		
		\textbf{Repeat}
		
		\textbf{~~Repeat}
		
		\textbf{\small{}\,\,\,\,\,\,\,\,\,\,-~}{\small{}Update
		$\eta=\eta^{\star}$,$\bm{\phi}=\bm{\phi}^{\star}$,$\bm{\lambda}=\bm{\lambda}^{\star}$} as defined in Section III-A.

		\textbf{\small{}\,\,\,\,\,\,\,\,\,\,-~}{\small{}Update $\bm{p}$ by
		\eqref{updateP}}.

		\textbf{\small{}\,\,\,\,\,\,\,\,\,\,-~}{\small{}Update $\bm{b}$ by
		\eqref{Update_d} }.

		\textbf{\small{}\,\,\,\,\,\,\,\,\,\,-~}{\small{}Update $\bm{g}$,
		$\hat{\bm{g}}$, $\bm{u}$, $\bm{d}$ by checking the first-order optimality}

\textbf{\small{}\,\,\,\,\,\,\,\,\,\,~~}{\small{}condition in turns}.

		\textbf{~~Until }{\small{}the value of \eqref{AL_obj} converges or reaching the max inner}

		\textbf{~~~~~~~~~}{\small{}iteration number $V$. Otherwise, let $v\leftarrow v+1$}.

		\textbf{\small{}\,\,\,\,\,\,\,\,\,\,-~If}{\small{}
			$\epsilon(\hat{\bm{z}}^{t+1})\leq \mu^t$, then update dual variables by
		\eqref{dualVar}},
		
		\textbf{\small{}\,\,\,\,\,\,\,\,\,\,-~Else} set{\small{}
			$\rho^{t+1}=\chi\rho^{t}$ end}.
		
		\textbf{\small{}\,\,\,\,\,\,\,\,\,\,-~Set}{\small{}
			$\mu^{t+1}=\chi\epsilon(\hat{\bm{z}}^{t+1})$.}
		
		\textbf{Until }{\small{}$\epsilon(\hat{\bm{z}}^{t+1})\leq \tau$ or $t\geq T$.
		Otherwise, let $t\leftarrow t+1$}.
	
	\end{algorithm}
	

	\section{SIMULATION RESULTS}
	Consider a single-cell system where $K=12$ users are uniformly distributed with a radius $r=200$ m. We set $N=64$, $S=12$, $M=8$, $\bar{b} = 3$, $L_p=10$, $P_B=200$ mW, $P_R=40$ mW, $P_S=5$ mW, $P_L=20$ mW, $\varphi=9\times10^{-12}$, $f_s=1$ GHz, $P_k^{\mathrm{max}}=10$ dBm, $\sigma^2_w=-100$ dBm, SNR = 10 dB, $b^{\min} = 1$, and $b^{\max} = 8$. The hyper-parameters of Algorithm 1 are chosen as: $\vartheta=10$, $\rho^0=10$, $\chi=0.7$, $\tau=10^{-4}$, $V=30$ and $T=150$. We adopt a discrete Fourier transformation (DFT) matrix as the codebook. The channel vector between the BS and the $k$-th user can be expressed as $\bm{h}_k=\sqrt{\frac{\mathrm{PL}_{k}N}{L}}\sum_{l=1}^{L}c_l^k \bm{a}(\phi_l^k), \forall k \in K$, where $L$ is the number of distinguishable paths, $c_l^k \sim \mathcal{CN}(0,1)$ and $\phi_l^k$ respectively are the complex gain and angle of arrival for the $l$-th path of user $k$, $\bm{a}(\phi)=\frac{1}{N}[1,e^{j\pi sin(\phi)},\dots,e^{j\pi (N-1) sin(\phi)}]$ is the receive array response vector, and $\textrm{PL}_{k}[\text{dB}] = 72 +29.2\log_{10}d_k + \xi$ is the large-scale fading gain between the BS and user $k$, where $d_k$ is the distance between the BS and user $k$ in meters, and $\xi \sim \mathcal{CN}(0,1)$ is the log-normal shadowing. 

	Besides the FDC and MMSQE-BA schemes, the following three schemes with fixed number of quantization bits for each ADC are also considered for comparison:
	\begin{itemize}
		\item \emph{RHC}: This is codebook-based  hybrid combining scheme with random selected codewords for RF combiner, digital combiner is optimized by maximizing the system EE.
		\item \emph{PHC}: This is the variation version of \cite{KARYAN_arxiv_2019} implemented by phase shifters with the consideration of jointly optimizing the user power allocation and digital combiner.
		\item \emph{SHC}: This is codebook-based  hybrid combining scheme, where the hybrid combiner is obtained by maximizing the system EE.
	\end{itemize}



	\begin{figure}
		\begin{minipage}[t]{0.25\textwidth}
			\centering
			\includegraphics[height=3cm,width=4cm]{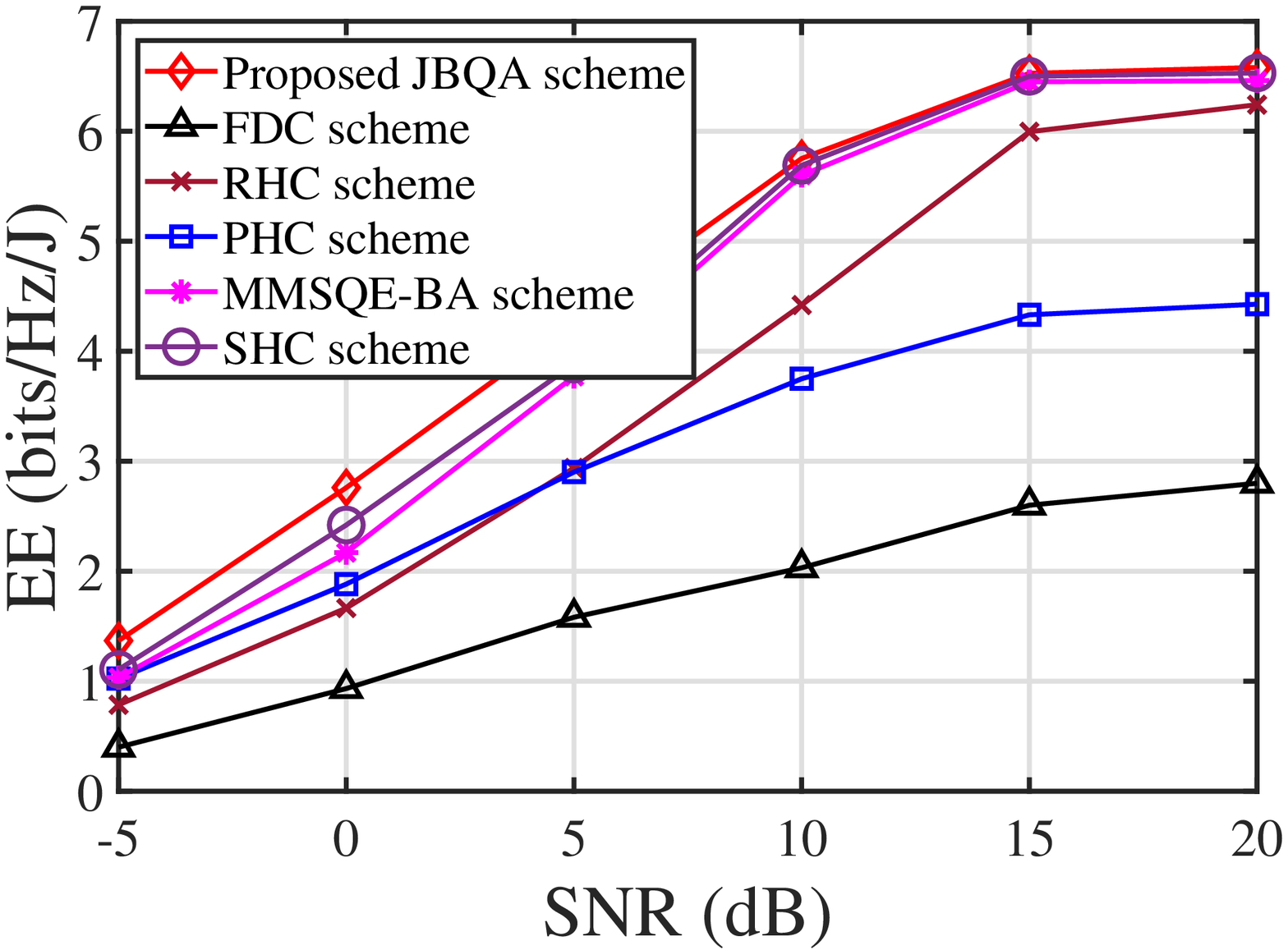}
			\caption{System EE versus SNR.}
			\label{fig:3_a}
		\end{minipage}%
		\begin{minipage}[t]{0.25\textwidth}
			\centering
			\includegraphics[height=3cm,width=4cm]{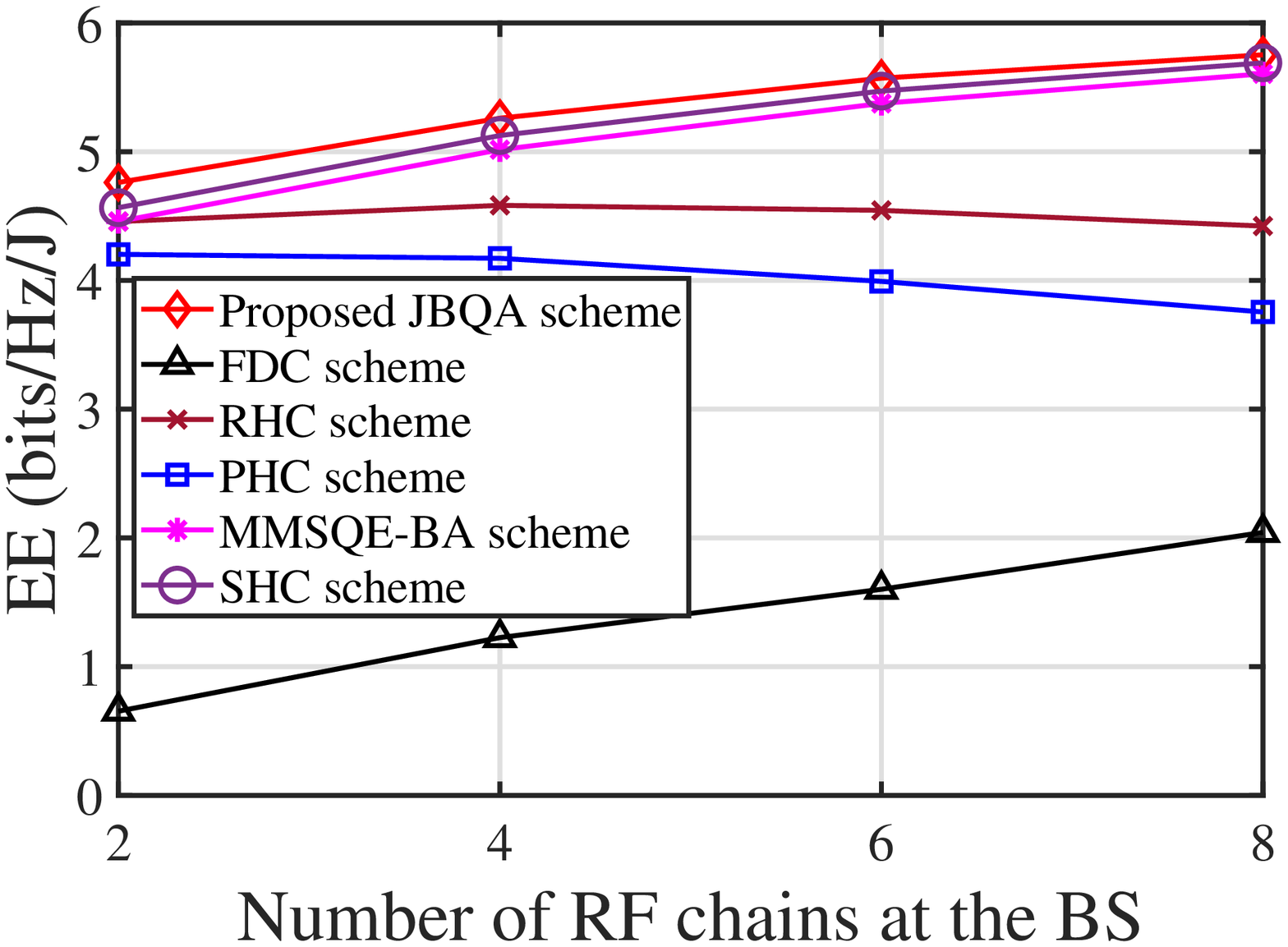}
			\caption{System EE versus M.}
			\label{fig:3_b}
		\end{minipage}
	\end{figure}

\begin{figure}
		\begin{minipage}[t]{0.25\textwidth}
			\centering
			\includegraphics[height=3cm,width=4cm]{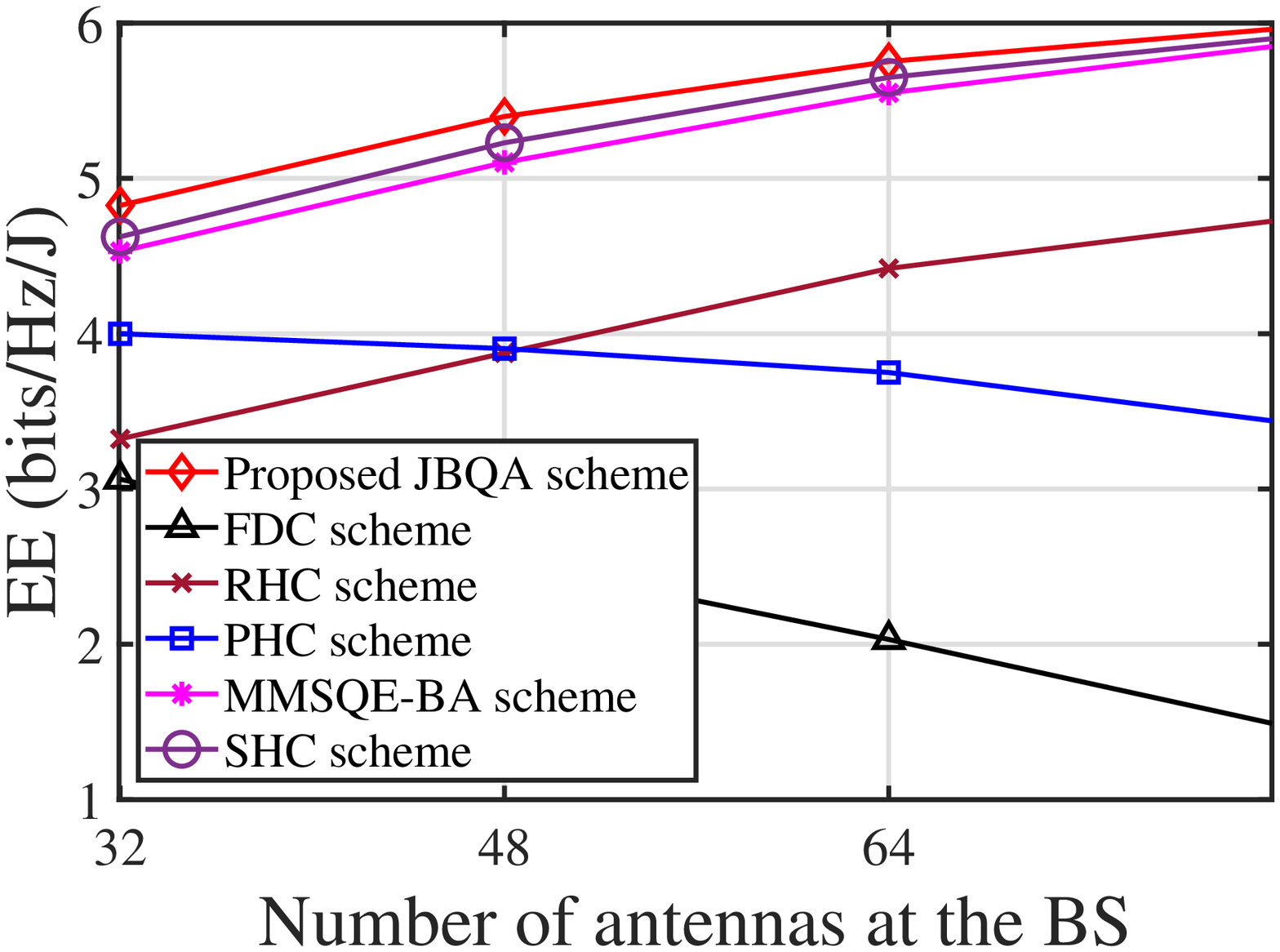}
			\caption{System EE versus N.}
			\label{fig:4_a}
		\end{minipage}%
		\begin{minipage}[t]{0.25\textwidth}
			\centering
			\includegraphics[height=3cm,width=4cm]{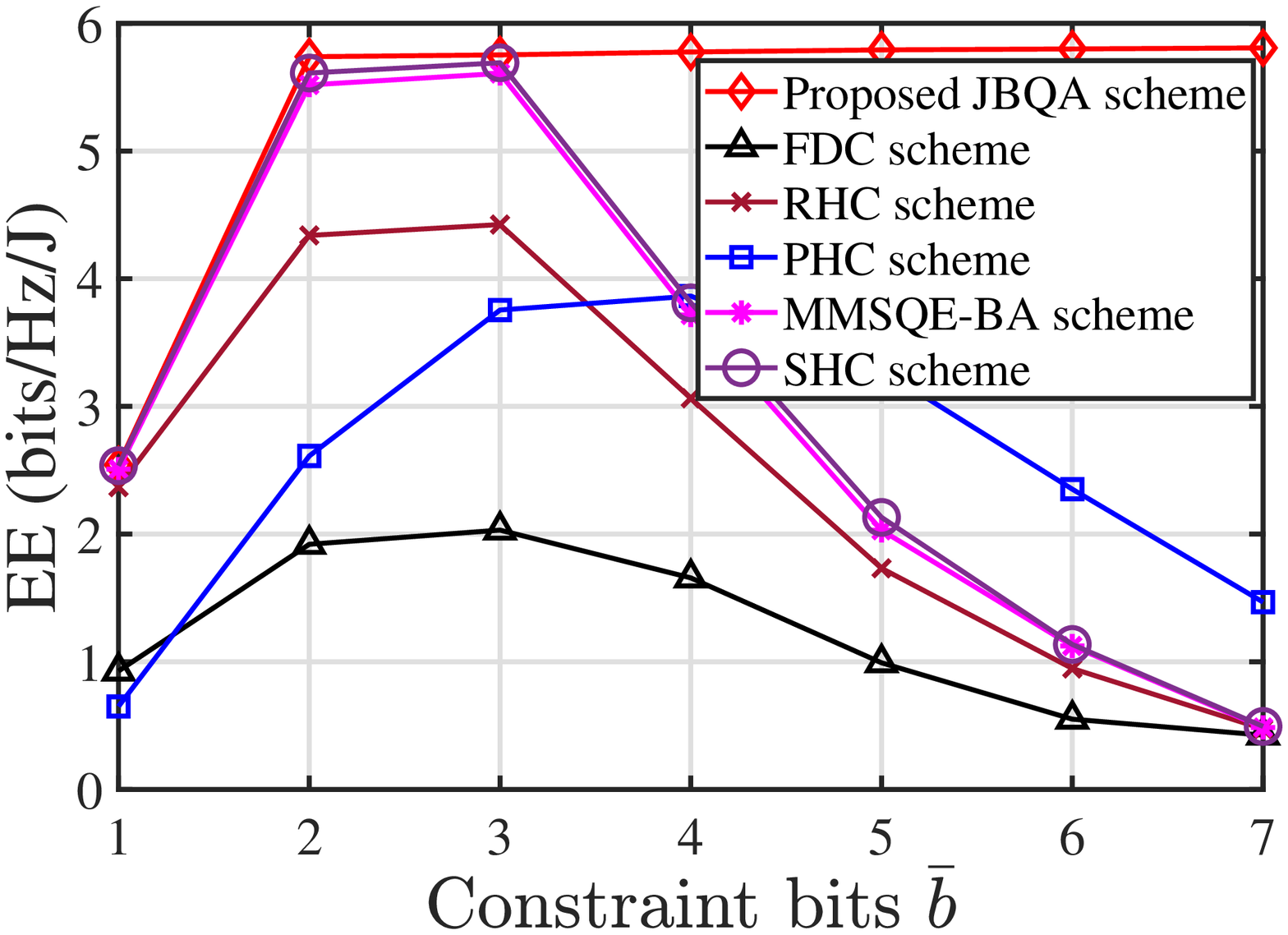}
			\caption{System EE versus $\bar{b}$.}
			\label{fig:4_b}
		\end{minipage}
	\end{figure}

	Fig. \ref{fig:3_a} plots the system EE versus the SNR for different schemes. It shows that the performance is monotonically increasing with SNR. Furthermore, we can see that the proposed JBQA scheme outperforms all the other competing schemes in all SNR regime, especially for the more practical moderate and low SNR regimes. The gap of performance becomes small in the high SNR regime. This is due to following reasons: 1) the hybrid combiner design based on the beamspace architecture is implemented with fewer RF chains and the reduced signal processing complexity as compared to the fully digital combiner and the phase-shifter-based hybrid combiner, and yet without notable performance degradation. 2) Unlike uniform quantization bit allocation, the proposed JBQA scheme would allocate more quantization bits to the ADC with a more favorable effective channel (the product of the channel and the RF combiner). However, when the allowed quantization bits is small and SNR is high, the proposed JBQA scheme tends to uniformly allocate all the quantization bits.  In Fig. \ref{fig:3_b} and Fig. \ref{fig:4_a}, we plot the system EE versus the number of antennas and RF chains at the BS, respectively. As expected, the proposed JBQA scheme achieves better EE over all the other competing schemes. In a nutshell, our proposed JBQA scheme can strike a better trade-off between the system throughput and power consumption.

	In Fig. \ref{fig:4_b}, we plot the system EE versus the allowed average quantization bits for different schemes. It is observed that the system EE achieved by all the other competing schemes increases first and then rapidly decreases after $\bar{b}=3$, while the proposed JBQA scheme keeps monotonically growing and saturate. As the number of average quantization bits increases, the power consumption of the RADCs increases exponentially. When the number of average quantization bits is larger than 3 bits, the EE of the system gradually tends to be saturate. From the perspective of maximizing the system EE, there is no need to deploy too excessive quantization bits at the ADCs.

	\section{CONCLUSION}
	In this work, we consider a multi-user mmWave uplink system, where the BS is equipped with a massive MIMO array and RADCs. We advocate a JBQA scheme to achieve high system throughput with the reduced hardware cost and power consumption. We formulate the optimization of the proposed JBQA scheme as a system EE maximization problem subject to some practical constraints. By adopting a series of transformations, we first recast this problem into a form amenable to optimization and then develop an efficient iterative algorithm for its solution based on PDD and MM methods. Simulations verify that the proposed JBQA scheme achieves significant gain over existing schemes.

\end{document}